\documentclass[12pt, twocolumn]{iopart}
\usepackage{graphicx}
\usepackage{subfig}
\begin{document}

\title[Computing local pressure]{Computing the local pressure in molecular dynamics simulations}

\author{Thomas W. Lion and Rosalind J. Allen}
\address{SUPA, School of Physics and Astronomy, The University of Edinburgh, 
Mayfield Road, Edinburgh EH9 3JZ, UK.}

\ead{T.Lion-2@sms.ed.ac.uk}

\begin{abstract}
Computer simulations of inhomogeneous soft matter systems often require accurate methods for computing the local pressure. We present  a simple derivation, based on the 
virial relation, of two equivalent expressions for the  local (atomistic) pressure in a molecular dynamics simulation. One of these expressions, previously derived by other authors 
via a different route, involves summation over interactions between particles within the region of interest; the other involves summation over interactions across the 
boundary of the region of interest. We illustrate our derivation  using simulations of a simple osmotic system; both expressions produce accurate results even when the region of interest over which the pressure is measured is very small.
%test these expressions 
%by measuring the local pressure in a homogeneous fluid and in an osmotic system and show that they produce accurate results even when the region of interest over which the pressure is measured is very small. 
\end{abstract}

%Uncomment for PACS numbers title message

\section{Introduction}
Molecular dynamics (MD) simulations are widely used to study spatially inhomogeneous soft matter systems. In such simulations, it is often necessary to compute the local pressure in a small region of the simulation box, containing only a few atoms or molecules. Examples include  calculations of interfacial free energies \cite{binder}, measurements of osmotic pressure gradients \cite{biben,inprep} and tests of coarse-grained hydrodynamic theories \cite{Schindler2010}. While accurate expressions for the local pressure exist, their derivation is rather involved. In this paper, we present a much simpler 
derivation, which leads to two equivalent expressions for the local pressure.  One of these expressions 
is analogous to the local stress tensor of Lutsko \cite{Lutsko1988} and  Cormier {\em{et al.}} \cite{Cormier2001}; 
the other is, to our knowledge, new, but is similar in spirit to the ``Method of Planes'' of Irving and Kirkwood \cite{IrvKirkPres} and Todd {\em{et al.}} \cite{ToddEvansMOP}. We show that both 
these expressions give accurate results for the local pressure in soft matter systems,  even when computed over very small spatial regions.

\section{Background}

For a spatially homogeneous, closed system, the pressure is commonly computed by taking the average of an instantaneous ``pressure function'':
\begin{equation}
P = \langle {\mathcal{ P}} \rangle = \left\langle \frac{Nk_{B}T}{V} + \frac{1}{3V} \sum_{i=1}^{N-1}\sum_{j>i} \vec r_{ij}\cdot\vec f_{ij} \right\rangle
 \label{eq:virialp}	
\end{equation}
where $N$, $V$ and $T$ are the number of particles, volume and temperature, $k_B$ is Boltzmann's constant, $\vec r_{i}$ and $\vec r_{j}$ are the positions of
 particles $i$ and $j$, $\vec r_{ij} \equiv \vec r_{i}-\vec r_{j}$, and $\vec f_{ij}$ denotes the force exerted on particle $i$ by particle $j$   
\cite{Cheung1977,Allen1992}; the double sum runs over all pairs of particles, avoiding double counting. The first and second terms in Eq.(\ref{eq:virialp}) arise from the kinetic energy of the particles and from interparticle interactions, respectively. Expression (\ref{eq:virialp}) can be derived in a few steps starting from the Clausius 
virial relation
\begin{equation}
 \left\langle \sum_{i=1}^N \frac{ |\vec p_i|^2 }{m_i} + \sum_{i=1}^N  \vec r_{i} \cdot \vec F_{i} \right\rangle = 0,
 \label{eq:clausius}	
\end{equation}  
in which  $ \vec p_i$ and $m_i$ are the momentum and mass of particle $i$, and $\vec F_{i}$ is the total force acting 
on particle $i$, due to other particles and the walls of the container  \cite{clausiusvirial1870, ErpWood1977}.

\begin{figure}[h!]
\begin{center}
{\rotatebox{0}{{\includegraphics[scale=0.6]{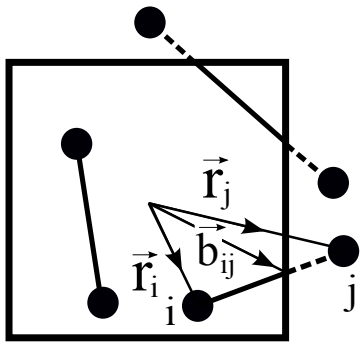}}}}
\caption{Contributions to the interaction part of Eq.(\ref{eq:CormierStress}) from different pairs of particles: the fraction $l_{ij}$ of the line joining particles $i$ and $j$ 
which lies inside the region of interest is shown as a solid line. $\vec b_{ij}$ denotes the position where the line joining particles $i$ and $j$ crosses the boundary.}
\label{fig:lijfraction}
\end{center}
\end{figure}

For spatially inhomogeneous systems, one can measure directly the pressure  across a local plane within the simulation box 
 via the ``method of planes'' (MOP), first proposed by Irving and Kirkwood \cite{IrvKirkPres} and later rederived by Todd {\em{et al.}} \cite{ToddEvansMOP}; this works well when the plane over which the pressure is computed is large, but leads to poor 
statistical sampling when computing the local pressure in a small region (i.e. over a small plane). 
Alternatively, the  pressure in a local region of interest can be 
measured using a local version of Eq.(\ref{eq:virialp}). This has the advantage that the  region of interest can (in principle) be of arbitrary shape, and that 
statistical averages are taken over a volume rather than an area.  The following local pressure function was proposed by Lutsko  \cite{Lutsko1988} and later 
reformulated by Cormier {\em{et al.}} \cite{Cormier2001} (note that these authors considered the full stress tensor, while, for simplicity, we focus only on the scalar 
pressure, which we assume to be locally isotropic):
\begin{equation}
P(\vec r) = \frac{1}{3\Omega}\left \langle \sum_{i=1}^{N}\frac{|p_{i}|^2}{m_i}\Lambda_i + \sum_{i=1}^{N-1} \sum_{j > i} (\vec f_{ij} \cdot \vec r_{ij}) l_{ij} 
\right \rangle .
\label{eq:CormierStress}
\end{equation}
Here, $\Omega$ is the volume of the region of interest, centred on  $\vec r$, $\Lambda_i$ is unity if particle $i$ lies within the volume $\Omega$, and zero otherwise, 
and  $l_{ij}$ is the fraction ($0\le l_{ij} \le 1$) of the line joining particles $i$ and $j$ that lies within $\Omega$ (see Figure \ref{fig:lijfraction}). The local pressure expression (\ref{eq:CormierStress}) is analogous to the 
global one (\ref{eq:virialp}), but with two important differences. Firstly, in the kinetic part of 
Eq.(\ref{eq:CormierStress}), only those particles which are inside the region of interest (at time $t$) are included. Secondly, the components of the interaction term are 
weighted by the fraction of the line joining particles $i$ and $j$ that is inside the region of interest; this highlights the crucial importance of correctly accounting for interparticle interactions which cross the boundary of the region of interest. Note that particles $i$ and $j$ may both be outside the 
region yet still contribute to the interaction pressure; see Figure \ref{fig:lijfraction}. 

Both Lutsko \cite{Lutsko1988} and Cormier {\em{et al.}} \cite{Cormier2001} derive the stress tensor version of  Eq.(\ref{eq:CormierStress}) by Fourier transforming the 
continuity equation for the local momentum flux and applying Newton's second law. In this paper, we present a simpler and arguably more intuitive, real space 
derivation, which is directly analogous to the derivation of the global pressure expression, Eq.(\ref{eq:virialp}), from the Clausius virial relation (\ref{eq:clausius}). 
%Our approach makes use of Schweitz's virial relation for open systems \cite{Schweitz1977}. 
This derivation leads  both to Eq.(\ref{eq:CormierStress}) and 
to a new expression for the local pressure involving the flux of particle momentum across the boundaries of the region of interest, together with cross-boundary 
interactions. The equivalence between these two expressions shows directly how the relation between surface flux and volume pressure measurements extends to the atomic scale. Our approach may also prove useful in future for deriving local pressure expressions for systems with different dynamical rules, such as run-and-tumble swimmers \cite{cates} or particles with viscous dynamics \cite{Allen1992}.

%We  use molecular dynamics simulations, first of a homogeneous fluid and then of an osmotic system,  to show that these 
%relations provide an accurate measurement of the local pressure, even when the local region of interest is very small. 

\section{Derivation of expressions for the local pressure}
\label{sec:deriv}

To compute the local pressure at some position $\vec r$ in the system,  we consider a local region of fluid, of volume $\Omega$, centred on $\vec r$. Particles within this region interact with other particles both inside and outside the 
region. During a given time interval, particles will enter and leave the region of interest.

\subsection{The Schweitz virial relation}
We begin with an analogue of the Clausius virial relation (\ref{eq:clausius}), derived by Schweitz, for open systems \cite{Schweitz1977}. The Schweitz virial relation states that
\begin{equation}
 \left\langle \sum_{i=1}^N \frac{ |\vec p_i|^2 }{m_i} \Lambda_i(\vec r ) + \sum_{i=1}^N\vec r_{i} \cdot \left(  \sum_{j \ne i} \vec f_{ij} \right) 
\Lambda_i(\vec r ) + \sum_{i=1}^N \left(\vec r_i \cdot \vec p_i\right) \dot{\Lambda_i} \right\rangle = 0,
 \label{eq:schweitz}	
\end{equation}
where, as above,  the function  $\Lambda_i(t)$ measures whether or not  particle $i$ is within the region of interest at 
time $t$, and its time derivative $d\Lambda_i(t)/dt \equiv \dot{\Lambda_i}$ produces a positive or negative $\delta$-function peak at the moment when 
particle $i$ enters or leaves the region of interest \footnote{For closed systems, the Clausius virial relation, Eq.(\ref{eq:clausius}), can be derived by setting the time derivative of the function  $G = \left \langle \sum_{i=1}^N \vec r_i \cdot \vec p_i \right \rangle$ to zero in steady state. For open systems, the derivation of the Schweitz virial relation follows a similar route, but takes into  account the contributions to $G_{local}(\vec r) = \left \langle \sum_{i=1}^N (\vec r_i \cdot \vec p_i) \Lambda_i(\vec r ) \right \rangle$ from particles entering and leaving the system \cite{Schweitz1977}.}. The first term in Eq.(\ref{eq:schweitz}), $\left\langle \sum_{i=1}^N \frac{ |\vec p_i|^2 }{m_i} \Lambda_i(\vec r ) \right \rangle \equiv E_{kin}$, is the average kinetic energy of particles in the region of interest and is directly analogous to the first term in Eq.(\ref{eq:clausius}). The second, interaction, term is  analogous to the second term in Eq.(\ref{eq:clausius}) -- we assume that there 
are no external forces so the total force on particle $i$ is given by the sum of interactions with all other particles in the system.  The final term in 
Eq.(\ref{eq:schweitz}), $\left \langle  \sum_{i=1}^N \left(\vec r_i \cdot \vec p_i\right) \dot{\Lambda_i} \right\rangle \equiv \Phi$,  accounts for the exchange of particles between the region of interest and its surroundings. Particles entering the region contribute $\vec r_i \cdot \vec p_i$  while those leaving contribute  $-\vec r_i \cdot \vec p_i$; these do not cancel because  the momentum vectors $\vec p_i$ for particles entering and leaving are  opposite in sign.

We now split the 
second term in Eq.(\ref{eq:schweitz}) into  contributions due to interactions with particles inside and outside the region of interest:
\begin{equation}
\sum_{i=1}^N\vec r_{i} \cdot \left(  \sum_{j \ne i} \vec f_{ij} \right) 
\Lambda_i(\vec r ) = \mathcal{V}_{int} +  \mathcal{V}_{ext},
\label{eq:split}
\end{equation}
where  $\mathcal{V}_{int}  \equiv \left \langle \sum_{i=1}^{N} \sum_{j\ne i} \vec r_{i} \cdot \vec f_{ij}\,\, \Lambda_i\Lambda_j \right\rangle = \left \langle \sum_{i=1}^{N-1} \sum_{j > i} \vec r_{ij} \cdot \vec f_{ij}\,\, \Lambda_i\Lambda_j \right\rangle \equiv \left \langle \sum_{\mathrm{in\,in}}  \vec r_{ij} \cdot \vec f_{ij}\right\rangle $\footnote{here we have used the fact that $\sum_{i=1}^N \sum_{j \ne i} \vec r_{i} \cdot \vec f_{ij} = \sum_{i=1}^{N-1} \sum_{j>i}\vec r_{ij} \cdot \vec f_{ij} $, since $ \vec f_{ij}=- \vec f_{ji}$ \cite{Allen1992}. We have also introduced a new notation: $\sum_{\mathrm{in\,in}} g_{ij} = \sum_{i=1}^{N-1} \sum_{j > i} \Lambda_i\Lambda_j g_{ij}$. Likewise, we define $\sum_{\mathrm{out\,out}} g_{ij} = \sum_{i=1}^{N-1} \sum_{j > i} (1-\Lambda_i)(1-\Lambda_j) g_{ij}$ and $\sum_{\mathrm{in\,out}} g_{ij} = \sum_{i=1}^{N} \sum_{j \ne i} \Lambda_i(1-\Lambda_j) g_{ij} = \sum_{i=1}^{N-1} \sum_{j > i} [\Lambda_i(1-\Lambda_j) g_{ij} + \Lambda_j(1-\Lambda_i) g_{ji}]$.} contains contributions where both particles are inside the region of interest and $\mathcal{V}_{ext}  \equiv \left \langle \sum_{i=1}^{N} \sum_{j\ne i}  \vec r_{i} \cdot \vec f_{ij} \,\, \Lambda_i(1-\Lambda_j) \right\rangle  \equiv \left \langle \sum_{\mathrm{in\,out}} \vec r_{i} \cdot \vec f_{ij}\right\rangle$ contains contributions where particle $i$ is inside the region and particle $j$ is outside. Substituting Eq.(\ref{eq:split}) into Eq.(\ref{eq:schweitz}) allows us to write the Schweitz virial relation as
\begin{equation}
 E_{kin} + \mathcal{V}_{int} +  \mathcal{V}_{ext} + \Phi = 0 .
 \label{eq:schweitz2}	
\end{equation}

\subsection{Expressions for  the local pressure}
We next use the Schweitz virial relation to derive expressions for the local pressure. The local pressure has two components: a kinetic component, $P_{kin}(\vec r)$, which is given 
by the normal flux of particle momentum across the boundaries of the region of interest  
and an interaction component, $P_{int}(\vec r)$, which is the surface density of interparticle forces across the boundary. Throughout this paper  the normal to the boundary is assumed to point in the outward direction.

\subsubsection*{Kinetic component} 

The kinetic component of the local pressure can be related to the  component $\Phi \equiv \left \langle \sum_{i=1}^N \left(\vec r_i \cdot 
\vec p_i\right) \dot{\Lambda_i} \right\rangle$ of the Schweitz virial relation. To see this, we split the particle momentum, $\vec p_i$,  into its components normal and tangential to the boundary: 
$ \vec p_i =   (\vec p_i \cdot \hat n)\hat n + (\vec p_i \cdot \hat t)\hat t $, to give
\begin{equation}\label{eq:kin1}
\Phi = \left \langle \sum_{i=1}^N  \dot{\Lambda_i} \left[ (\vec p_i \cdot \hat n) (\vec r_i \cdot \hat n) + (\vec p_i \cdot \hat t)(\vec r_i \cdot 
\hat t )\right]\right\rangle .
\end{equation}
Assuming that the density of particles is uniform across the region of interest, the second term in Eq.(\ref{eq:kin1}) averages to zero. Next, we notice that 
because $\dot{\Lambda_i}$ is nonzero only when a particle is at the boundary, $\vec r_{i}\cdot \hat{n}$ may be taken outside the summation. Assuming 
(without loss of generality) that the region of interest is cubic with the origin at its centre and sides of length $L$,  $\vec r_{i}\cdot \hat{n}=L/2$, 
and the total (outward) momentum flux $\left\langle\sum_i (\vec p_i \cdot \hat n)\dot{\Lambda_i}\right\rangle$ across each of the 6 faces is $-P_{kin} L^2$. Eq.(\ref{eq:kin1}) therefore reduces to
\begin{equation}\label{eq:kin2}
\Phi = -3 \Omega P_{kin}(\vec r).
\end{equation}

\subsubsection*{Interaction component}
In a similar way, the interaction component of the local pressure, $P_{int}(\vec r)$, can be related to the component $\mathcal{V}_{ext}$ of the Schweitz virial relation.  We first note that the position  vector $\vec r_{i}$  may be written as $\vec r_{i} = \vec b_{ij} + \vec r_{ij} l_{ij}$, 
where $\vec b_{ij}$ denotes the position where the line linking particles $i$ and $j$ crosses the boundary of the region of interest, and  $l_{ij}$ is the fraction of the line linking particles $i$ and $j$ that lies inside the region of interest  (see Figure \ref{fig:lijfraction}).  $\mathcal{V}_{ext}$ is then given by
\begin{eqnarray} \label{eq:schweitz2c}	
\mathcal{V}_{ext}  &=&  \left \langle  \sum_{\mathrm{in\,out}} \vec b_{ij} \cdot \vec f_{ij} +  l_{ij}\left(\vec r_{ij} \cdot \vec f_{ij}\right)  
\right \rangle\\\nonumber &=& \left \langle  \sum_{\mathrm{in\,out}} (\vec f_{ij} \cdot \hat n)\left(\vec b_{ij} \cdot \hat n\right)\right \rangle + C_t  + \left \langle \sum_{\mathrm{in\,out}} l_{ij}\left(\vec r_{ij} \cdot \vec f_{ij}\right)
\right \rangle,
\end{eqnarray}
where the second line follows from splitting the  interparticle force $\vec f_{ij}$ into its components normal and tangential to the boundary: 
$\vec f_{ij} =   (\vec f_{ij} \cdot \hat n)\hat n + (\vec f_{ij} \cdot \hat t)\hat t$, and 
$C_t=\left \langle \sum_{\mathrm{in\,out}} (\vec f_{ij} \cdot \hat t)\left(\vec b_{ij} \cdot \hat t\right)\right \rangle $. 
Focusing on the first term, we note that $\vec b_{ij}$ points to the boundary and so (assuming the same cubic geometry as above), $\vec b_{ij} \cdot \hat n = L/2$. 
Denoting as $\sigma $ the average outward normal force per unit area crossing the boundary, due to the $(i\,{\mathrm{in}},j\,{\mathrm{out}})$ interactions, we obtain
\begin{eqnarray}\label{eq:schweitzBoundary}
\left \langle \sum_{\mathrm{in\,out}} (\vec f_{ij} \cdot \hat n)\left(\vec b_{ij} \cdot \hat n\right)\right \rangle =  -6 \left(\frac{L}{2}\right)(L^2 \sigma) = -3 \Omega  \sigma .
\end{eqnarray}
The interaction component of the pressure, $P_{int}(\vec r)$, is equal to the {\em{total}} surface density of normal force crossing the boundary: $P_{int} = \sigma + \xi$, 
where $\xi$ is the normal contribution  due to pairs of particles $i$ and $j$ which are both outside the region of interest (see Figure \ref{fig:lijfraction}). 
We demonstrate in the Appendix that 

\begin{equation}\label{eq:xi}
-3\Omega \xi =  C_t  - \left \langle \sum_{\mathrm{out\,out}} l_{ij}\left(\vec r_{ij} \cdot \vec f_{ij}\right)\right \rangle
\end{equation}
so that, putting together Eqs(\ref{eq:schweitz2c})-(\ref{eq:xi}), we can relate $P_{int}$ to $\mathcal{V}_{ext}$ by:
\begin{eqnarray}
-3\Omega P_{int}(\vec r) &=& -3\Omega (\sigma + \xi) =  \mathcal{V}_{ext} -  \mathcal{V}_{corr},
 \label{eq:schweitz3}	
\end{eqnarray}
where
\begin{equation}
 \mathcal{V}_{corr} \equiv \left \langle \sum_{\mathrm{in\,out}}  l_{ij}\left(\vec r_{ij} \cdot \vec f_{ij} \right) + \sum_{\mathrm{out\,out}}  l_{ij}\left(\vec r_{ij} \cdot \vec f_{ij} \right)\right \rangle.
\end{equation}

\subsubsection*{``Boundary'' expression for the local pressure} 
An expression for the local pressure, $P(\vec r)$, can be obtained by combining Eqs.(\ref{eq:kin2}) and (\ref{eq:schweitz3}):
\begin{eqnarray}\label{eq:res1}
P(\vec r) &=& P_{kin} + P_{int} =  -\frac{1}{3\Omega} \left[\Phi +  \mathcal{V}_{ext}-\mathcal{V}_{corr} \right]\\\nonumber &=& 
 -\frac{1}{3\Omega}\left[\left \langle \sum_{i=1}^N \left(\vec r_i \cdot \vec p_i\right) \dot{\Lambda_i} +  \sum_{\mathrm{in\,out}}\left(\vec r_i  - l_{ij} \vec r_{ij} \right) \cdot \vec f_{ij} - \sum_{\mathrm{out\,out}} l_{ij} \vec r_{ij} \cdot \vec f_{ij} \right \rangle\right].
\end{eqnarray}
Eq.(\ref{eq:res1}) provides a simple prescription for computing the local pressure. The first term sums over all particles which enter or leave the region of interest and 
is equivalent to the momentum flux density due to particles crossing the boundary, 
while the remaining terms, which account for the force density at the boundary 
due to interparticle interactions, sum over all pairs of particles for which the line connecting the two particles crosses the boundary of the region of interest. In the case where the region of interest is large, $ P(\vec r)$ is dominated by the contributions of $\Phi$ and $\mathcal{V}_{ext}$ ($\mathcal{V}_{corr}$ becomes negligible); however, as we show below in Figure \ref{fig:OpenSystem}, $\mathcal{V}_{corr}$  makes an important contribution when the region of interest is small. Eq.(\ref{eq:res1}) provides an alternative to existing local pressure expressions, and demonstrates explicitly how the relation between surface flux and volume pressure expressions extends to very small regions of space.

\subsection*{``Volume'' expression for the local pressure}
The Schweitz virial relation provides a direct route from this ``boundary'' expression to the more usual expression for the local pressure, which involves   a sum over particles {\em{within}} the region of interest. Inserting Eq.(\ref{eq:res1}) into the Schweitz  relation  (\ref{eq:schweitz2}), we obtain:
\begin{eqnarray}\label{eq:res2}
&& P(\vec r)  =  \frac{1}{3\Omega} \left[E_{kin} +  \mathcal{V}_{int}+\mathcal{V}_{corr} \right]\\\nonumber &=&  
\frac{1}{3\Omega}\left \langle \sum_{i=1}^N \frac{ |\vec p_i|^2 }{m_i} \Lambda_i(\vec r ) + \sum_{\mathrm{in\,in}}  \vec r_{ij} \cdot \vec f_{ij} + \sum_{\mathrm{in\,out}}  l_{ij}\left(\vec r_{ij} \cdot \vec f_{ij} \right) + \sum_{\mathrm{out\,out}}  l_{ij}\left(\vec r_{ij} \cdot \vec f_{ij} \right) \right \rangle\\\nonumber &=&  
\frac{1}{3\Omega}\left \langle \sum_{i=1}^N \frac{ |\vec p_i|^2 }{m_i} \Lambda_i(\vec r ) + \sum_{i}
\sum_{j>i} l_{ij}\left( \vec r_{ij} \cdot \vec f_{ij}\right)  \right \rangle,
\end{eqnarray}
where the last line follows from the fact that $l_{ij}=1$ if both particles are inside the region of interest (assuming the boundary is everywhere concave). 
Eq.(\ref{eq:res2}) is identical to Eq.(\ref{eq:CormierStress}), and constitutes a local version of the global instantaneous pressure function, Eq.(\ref{eq:virialp}).

\section{Molecular dynamics simulations}\label{sec:md}

We now illustrate, using molecular dynamics simulations, the Schweitz virial relation (\ref{eq:schweitz2}) as well as the two expressions for the local pressure, 
Eqs.(\ref{eq:res1}) and (\ref{eq:res2}). In our simulations, a periodic box contains 5000 particles at a density $\rho = 0.8$ (reduced units \cite{Allen1992}), which interact 
via a repulsive Weeks, Chandler, Andersen (WCA) potential \cite{WCA1971} (a truncated and shifted Lennard-Jones potential). The particle size $\sigma =1$, the interaction parameter $\epsilon=1$ (both in reduced units) and the box length is $18.42\sigma$.  The system is simulated using the velocity Verlet algorithm \cite{Allen1992} with timestep $\Delta t = 0.001$ (reduced units) and is  maintained in the canonical (NVT) ensemble using a Nos\'{e}-Hoover thermostat at temperature, $T = 1.0$ (reduced units). All  runs are equilibrated for $4 \times 10^4$ timesteps prior to data collection, and data is collected over at least $2 \times 10^6$ timesteps. Errors are computed using bootstrapping \cite{Bootstrap}, with 1000 replica datasets. 
%, where each distribution comprises 1000 averages. In the homogeneous fluid, data is taken from over 5 million time steps ($\Delta t = 0.001$) of equilibrated data. In the osmotic system we use at least $\sim$ 2 million time steps of equlibrated data.

\subsection{A homogeneous fluid}
We first consider a homogeneous fluid, for which the local pressure, measured using Eqs.(\ref{eq:res1}) and (\ref{eq:res2}), should match the global pressure, measured using Eq.(\ref{eq:virialp}). We define a cubic region of interest, located in the centre of our simulation box, whose size $L$ we vary from $L=5.5\sigma$ to $L=3\sigma$. For this smallest value of $L$, the region of interest contains only $\sim 21$ particles on average.   

\begin{figure}[h!]
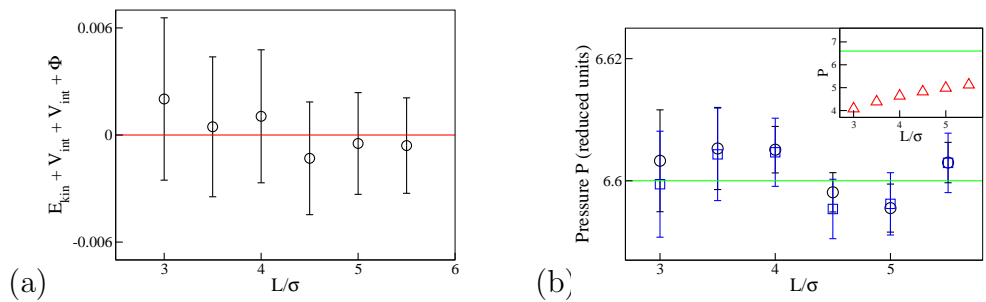

\begin{center}
\makebox[15pt][l]{(a)}{\rotatebox{0}{{\includegraphics[scale=0.22]{Lion_Allen_Fig2a.eps}}}}\hspace{1cm}\makebox[15pt][l]{(b)}{\rotatebox{0}{{\includegraphics[scale=0.22]{Lion_Allen_Fig2b.eps}}}}
\caption{Molecular dynamics simulation results for the homogeneous fluid. (a): $E_{kin}+\mathcal{V}_{int}+\mathcal{V}_{ext}+\Phi$ as a function of the size $L$ of 
the region of interest. (b): Local pressure in the region 
of interest, computed using  Eq.(\ref{eq:res1}) (blue squares) and  Eq.(\ref{eq:res2}) (black circles). 
The apparent correlation between these results arises because the same simulation data set was used in both cases.  The green line shows the global pressure computed  using Eq.(\ref{eq:virialp}). The inset shows the  pressure computed using Eq.(\ref{eq:res2}), not including $\mathcal{V}_{corr}$ (red triangles; note the very different scale on the pressure axis). \label{fig:OpenSystem}}
\end{center}
\end{figure}

Figure  \ref{fig:OpenSystem}a shows $E_{kin}+\mathcal{V}_{int}+\mathcal{V}_{ext}+\Phi$ as a function of the size $L$ of the region of interest. As predicted by 
the Schweitz virial relation (\ref{eq:schweitz2}), this quantity is zero within our error bars 
(note that of the individual terms in this sum, two are $\sim$ 5 and the other two are $\sim$ 0.8). Figure  \ref{fig:OpenSystem}b shows the local pressure $P(\vec r)$ 
in the region of interest, computed using expressions (\ref{eq:res1}) and (\ref{eq:res2}), as a function of  $L$. Both expressions give results in 
excellent agreement with the global pressure across the whole simulation box, using Eq.(\ref{eq:virialp}). The inset to Figure  \ref{fig:OpenSystem}b, which shows  Eq.(\ref{eq:res2}), neglecting the $\mathcal{V}_{corr}$ term, demonstrates 
 the importance of 
correctly accounting for the boundary terms: neglecting $\mathcal{V}_{corr}$ gives a large error, which increases as $L$ decreases.

\subsection{An osmotic system}

\begin{figure}[h!]
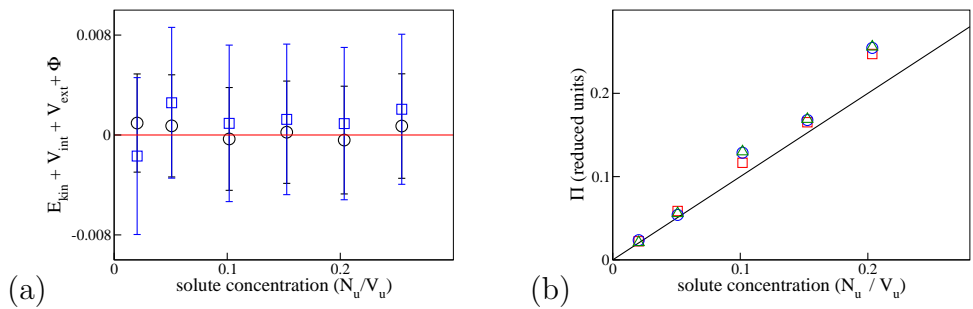

\begin{center}
\makebox[15pt][l]{(a)}{\rotatebox{0}{{\includegraphics[scale=0.22]{Lion_Allen_Fig3a.eps}}}}\hspace{1cm}\makebox[15pt][l]{(b)}{\rotatebox{0}{{\includegraphics[scale=0.22]{Lion_Allen_Fig3b.eps}}}}
\caption{Molecular Dynamics simulations for the osmotic system. (a): $E_{kin}+\mathcal{V}_{int}+\mathcal{V}_{ext}+\Phi$, as a function of the concentration $N_u/V_u$ of 
solute particles,  computed for a small subregion (of dimensions $L_{x} = L_{y} = L_{z} = 6\sigma$)  within the solution region (black circles) 
and for a small subregion ($L_{x} = 2\sigma$, $L_{y} = L_{z} = 7\sigma)$  outside the solution (blue squares). (b): Osmotic pressure difference $\Pi$ between the solution and its surroundings, 
computed using Eq.(\ref{eq:res1}) (green triangles) and  using Eq.(\ref{eq:res2}) (blue circles) (for the same subregions). Results 
are also shown for a direct computation of the normal solute-membrane force per unit area (red squares). Error bars are smaller than the symbols. The prediction of the  
van't Hoff equation,  $\Pi = (N_u k_BT)/V_u$, is also shown (black line).} 
\label{fig:osmotic}
\end{center}
\end{figure}

We next consider an osmotic system, in which the local pressure differs in different parts of the simulation box. To construct this system we label a subset $N_u$ 
of the particles ``solute'' and the remaining particles ``solvent''. Solute particles are confined to a cubic region of volume $V_u\approx 786 \sigma^3$ in the  
center of the 
simulation box by a smooth confining potential; solvent particles do not experience this potential and are free to move throughout the simulation box. 
The confining potential acts as a semi-permeable membrane, resulting in an osmotic pressure difference, $\Pi$, between the ``solution'' region where the solutes are 
confined and the rest of the simulation box. 
%The van't Hoff equation \cite{vantHoff1887} predicts that, for dilute solutions, $\Pi$ should increase linearly with the concentration, $N_u/V_u$,
%of solute particles in the solution compartment: $\Pi = (N_u k_BT)/V_u$. Since our particles interact repulsively, we expect the true osmotic pressure 
%to deviate from the prediction of van't Hoff already at quite low solute concentrations. 

To compute the osmotic pressure, we  define local regions of interest  inside and outside the solution compartment (for dimensions see the caption of Figure \ref{fig:osmotic}).  Figure \ref{fig:osmotic}a shows that the Schweitz virial relation (\ref{eq:schweitz2}) is obeyed in both of these local regions, over a range of solute 
concentrations. The pressure in the two local regions can be computed using Eqs. (\ref{eq:res1}) and (\ref{eq:res2}); subtracting the results gives  the osmotic pressure 
difference, $\Pi$.  Figure  \ref{fig:osmotic}b shows  $\Pi$, as a function of the concentration $N_u/V_u$ of solute particles in the solution compartment; both methods for computing the local pressure produce results in excellent agreement with a direct calculation of the osmotic pressure obtained by measuring  the average confining force on the solutes, per unit area of 
the confining box. 
%Computing the local pressure using either the boundary method Eq.(\ref{eq:res1}) or the volume method Eq.(\ref{eq:res2}) produces results in excellent agreement with the direct calculation. As expected, $\Pi$ exceeds the van't Hoff prediction by an amount that increases 
%with increasing solute concentration.

\section{Conclusions}\label{sec:conc}
Accurate methods for computing the local pressure are essential for simulating inhomogeneous soft matter systems. In this paper, we have derived two equivalent expressions for the local pressure in a molecular dynamics simulation. The ``boundary'' 
expression, Eq.(\ref{eq:res1}), is, to our knowledge, new. This expression involves summation over interactions between particles within and outside the local region of interest and is similar in spirit to the 
``method of planes'' approach of Irving and Kirkwood \cite{IrvKirkPres} and of Todd {\em{et al.}} \cite{ToddEvansMOP}. The ``volume'' expression Eq.(\ref{eq:res2}) is a local analogue 
of the function commonly used to compute the global pressure in homogeneous simulations; this involves summation over interactions between pairs of particles 
within the region of interest. This expression was previously derived via a Fourier transform method by Lutsko \cite{Lutsko1988} and by Cormier {\em{et al.}} 
\cite{Cormier2001}; our derivation, based on the  Schweitz virial relation, provides a simple real-space alternative. Importantly, both local pressure expressions take  careful account of interactions close to the boundary: this is crucial when the region of 
interest is of the order of the particle size.

% We have demonstrated the validity of both expressions for computing the local pressure even in very small spatial 
%regions using molecular dynamics simulations of a homogeneous fluid and an osmotic system. 

The derivation presented in this paper, via the Schweitz virial relation, demonstrates explicitly how the equivalence between surface flux and volume expressions for the pressure, familiar from macroscopic systems, plays out on very small (atomistic) lengthscales. This approach may also prove useful in future for deriving local pressure expressions in systems whose dynamics are more complex: for example   systems with viscous dynamics \cite{Allen1992}, or active matter systems in which particles are self-propelled and/or chemotactic \cite{cates}. Here,  the Fourier transform method of Lutsko and Cormier {\em{et al.}} \cite{Lutsko1988,Cormier2001} might prove challenging, but we hope that our real-space method should hold with only minor modifications.

Finally, we note that an important assumption made in this work is that the  local pressure is isotropic: we therefore derive expressions for the scalar pressure rather than the local 
pressure tensor, as in previous work \cite{Lutsko1988,Cormier2001,IrvKirkPres,ToddEvansMOP}. We believe that it should be possible to extend the present derivation 
to obtain the analogous expressions for  the pressure tensor,  via  a tensor analogue of the Schweitz virial relation. For the present, we leave this as an 
interesting avenue for future work.

\section*{Acknowledgments}
The authors thank Mike Cates, Daan Frenkel, Davide Marenduzzo and Juan Venegas-Ortiz for useful discussions. TL was supported by an EPSRC DTA studentship;  RJA was supported by a Royal Society University Research Fellowship.

\appendix

\section*{Appendix: Pairs of particles which are both outside the region of interest}
\label{app:Gij}
\begin{figure}[h!]
\begin{center}
\makebox[20pt][l]{(a)}{\rotatebox{0}{{\includegraphics[scale=0.6]{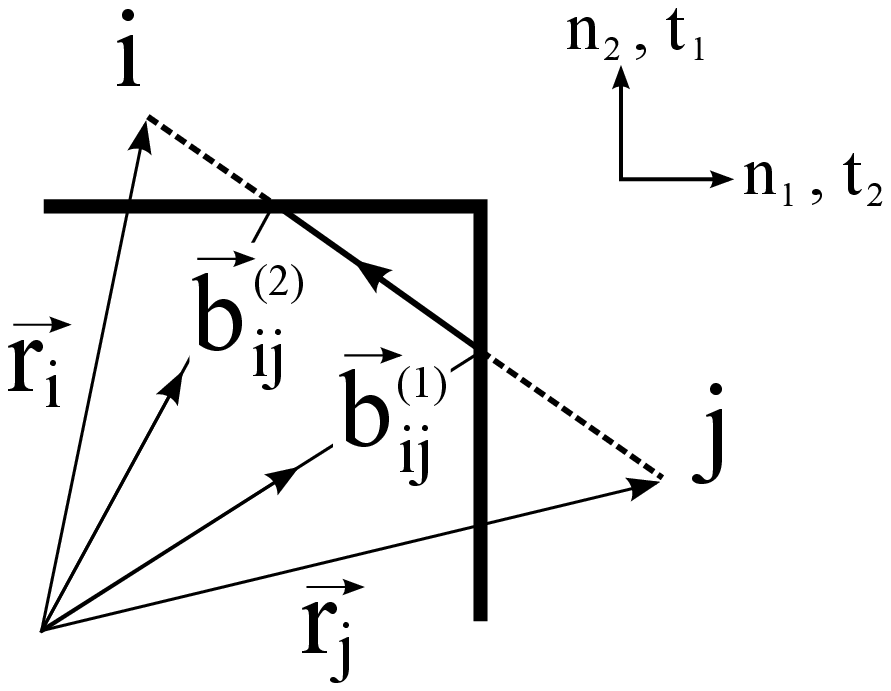}}}}\hspace{1cm}
\makebox[15pt][l]{(b)}{\rotatebox{0}{{\includegraphics[scale=0.45]{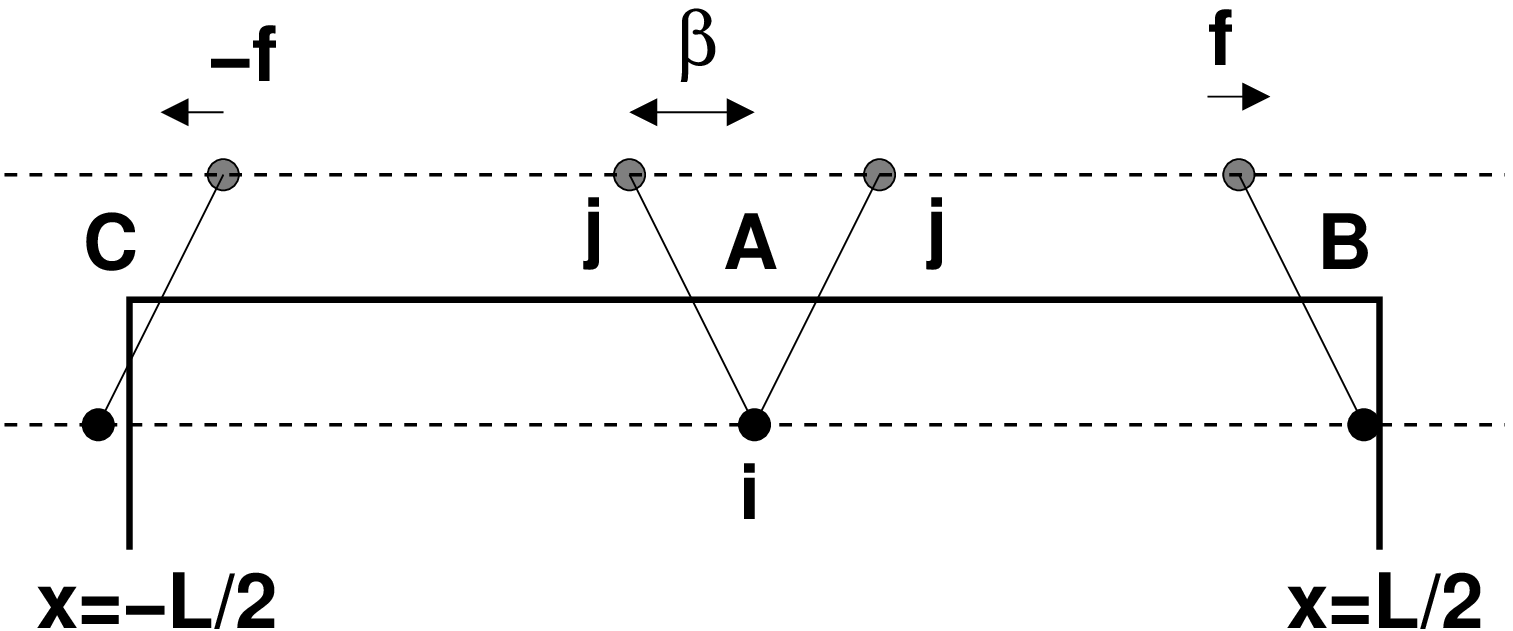}}}}
\caption{(a) Geometric illustration of the case where particles $i$ and $j$ are both outside the region of interest. (b) 
Illustration of the simplified one dimensional scenario used to show that $C_t=-D_t$. }
\label{fig:app}
\end{center}
\end{figure}

Here, we discuss the contribution to the interaction component of the pressure made by pairs of particles $i$ and $j$, in the cases where both particles are outside 
the region of interest, but (as illustrated in Figure \ref{fig:lijfraction}), the line joining particles $i$ and $j$ crosses the boundary. 
Our aim is to derive Eq.(\ref{eq:xi}). Assuming that the region of interest is cubic (with side length $L$ and origin at the centre), the line 
joining particles $i$ and $j$ crosses the boundary on two different faces. We define these crossing points by the position vectors $\vec b_{ij}^{(1)}$ and 
$\vec b_{ij}^{(2)}$. Figure  \ref{fig:app}a illustrates that $l_{ij} \vec r_{ij} = \vec b_{ij}^{(2)} - \vec b_{ij}^{(1)}$.  This allows us to write
\begin{equation}
 \left \langle \sum_{\mathrm{out\,out}} l_{ij}\left(\vec r_{ij} \cdot \vec f_{ij}\right)\right \rangle = \left \langle \sum_{\mathrm{out\,out}} \left( \vec b_{ij}^{(2)} \cdot \vec f_{ij} \right) - \left( \vec b_{ij}^{(1)}   \cdot \vec f_{ij}\right) \right \rangle.
\end{equation}
We now resolve $\vec b_{ij}^{(1)}$ and $\vec b_{ij}^{(2)}$ into components normal and tangential to the boundary on faces 1 and 2 respectively [$\vec b_{ij}^{(1)} =  (\vec b_{ij}^{(1)} \cdot \hat n^{(1)})\hat n^{(1)} + (\vec b_{ij}^{(1)} \cdot \hat t^{(1)})\hat t^{(1)}$ etc]. Noting that $(\vec b_{ij}^{(1)} \cdot \hat n^{(1)})=(\vec b_{ij}^{(2)} \cdot \hat n^{(2)})=L/2$,  we obtain
\begin{equation}\label{eq:app1}
 \left \langle \sum_{\mathrm{out\,out}} l_{ij}\left(\vec r_{ij} \cdot \vec f_{ij}\right)\right \rangle = 3 \Omega \xi - (D_t^{(1)}+ D_t^{(2)}),
\end{equation}
where $\xi$ is the (outward) normal force per unit area crossing the boundary due to particle pairs which are both outside the region of interest, $D_t^{(1)} = \left \langle \sum_{\mathrm{out\,out}}   (\vec b_{ij}^{(1)} \cdot \hat t^{(1)})(\vec f_{ij} \cdot \hat t^{(1)})\right \rangle$ and $D_t^{(2)} = \left \langle \sum_{\mathrm{out\,out}} (\vec b_{ij}^{(2)} \cdot \hat t^{(2)})(\vec f_{ji} \cdot \hat t^{(2)})\right \rangle$  (we use $\vec f_{ji} = -\vec f_{ij}$). Assuming that particles are homogeneously distributed across the region of interest, $D_t^{(1)}=D_t^{(2)} = D_t/2$ (by symmetry, see Figure \ref{fig:app}a).

We now demonstrate that $D_t = -C_t$, where $C_t$, as defined in Section \ref{sec:deriv}, is the tangential component of the ($i\,{\mathrm{in}},j\,{\mathrm{out}}$) 
contribution to $\mathcal{V}_{ext}$. We study the simplified  one dimensional scenario shown in Figure \ref{fig:app}b, in which we consider only particles $i$ and $j$ 
lying along two lines parallel to the $x$ axis. We further assume that interactions are restricted in range, such that  each particle $i$ interacts with only two 
interaction partners $j$, as shown in Figure  \ref{fig:app}b, with tangential forces $f$ in opposite directions. These interactions contribute respectively $f(x_i-\beta)$ 
and  $-f(x_i+\beta)$ to the sum $\sum_i \sum_j (\vec b_{ij} \cdot \hat t)(\vec f_{ij} \cdot \hat t)$. We now sum the contribution of the interactions shown in Figure  
\ref{fig:app}b for the cases where particle $i$ is inside the region of interest ($-L/2 < x_i < L/2$) and the line joining $i$ and $j$ crosses the top face of the 
region -- i.e. the contributions to $C_t$ for this face. The result is $f\beta d (\beta-L)$, where $d$ is the number of particles $i$ per unit length along the line 
\footnote{This result follows from noting that particles $i$ for which  ($-L/2+\beta < x_i < L/2-\beta$) have 2 partners and contribute $-2 f \beta$ (case A in Figure 
\ref{fig:app}b), particles for which $-L/2 < x_i < -L/2+\beta$ have 1 partner and contribute $-(x_i+\beta)f$ while particles for 
which $L/2-\beta < x_i < L/2$   have 1 partner and contribute $(x_i-\beta)f$ (case B in Figure \ref{fig:app}b); these contributions are then integrated over $-L/2 < x < L/2$: $df\left[-\int_{-L/2}^{-L/2+\beta}(x+\beta) dx - \int_{-L/2+\beta}^{L/2-\beta}2 \beta dx + \int_{L/2-\beta}^{L/2}(x-\beta) dx\right] = f\beta d(\beta-L)$.}. Next, we sum the 
contributions for the cases  where particle $i$ is {\em{outside}} the region of interest, but  the line joining $i$ and $j$ still crosses the top face of the region 
(i.e. the contributions to $D_t$ for this face). Contributions $\pm (x \mp \beta)f$ are made by particles $i$ for which 
$(-L/2-\beta) < x_i < -L/2$ or $L/2 < x_i < (L/2+\beta)$ (case C in Figure  \ref{fig:app}b). Integrating over these ranges of $x_i$, we obtain a total contribution to 
$D_t$ of $f\beta d(L-\beta)$. Thus the contributions to the 
($i\,{\mathrm{in}},j\,{\mathrm{out}}$) tangential term $C_t$ are exactly compensated by the contributions to the ($i\,{\mathrm{out}},j\,{\mathrm{out}}$) 
tangential term  $D_t$, and Eq.(\ref{eq:app1}) reduces to Eq.(\ref{eq:xi}).

\end{document}